\documentstyle[12pt]{article}
\makeatletter
\newdimen\normalarrayskip              
\newdimen\minarrayskip                 
\normalarrayskip\baselineskip
\minarrayskip\jot
\newif\ifold             \oldtrue            
\def\arraymode{\ifold\relax\else\displaystyle\fi} 
\def\eqnumphantom{\phantom{(\theequation)}}     
\def\@arrayskip{\ifold\baselineskip\z@\lineskip\z@
     \else
     \baselineskip\minarrayskip\lineskip2\minarrayskip\fi}
\def\@arrayclassz{\ifcase \@lastchclass \@acolampacol \or
\@ampacol \or \or \or \@addamp \or
   \@acolampacol \or \@firstampfalse \@acol \fi
\edef\@preamble{\@preamble
  \ifcase \@chnum
     \hfil$\relax\arraymode\@sharp$\hfil
     \or $\relax\arraymode\@sharp$\hfil
     \or \hfil$\relax\arraymode\@sharp$\fi}}
\def\@array[#1]#2{\setbox\@arstrutbox=\hbox{\vrule
     height\arraystretch \ht\strutbox
     depth\arraystretch \dp\strutbox
     width\z@}\@mkpream{#2}\edef\@preamble{\halign \noexpand\@halignto
\bgroup \tabskip\z@ \@arstrut \@preamble \tabskip\z@ \cr}%
\let\@startpbox\@@startpbox \let\@endpbox\@@endpbox
  \if #1t\vtop \else \if#1b\vbox \else \vcenter \fi\fi
  \bgroup \let\par\relax
  \let\@sharp##\let\protect\relax
  \@arrayskip\@preamble}
\def\eqnarray{\stepcounter{equation}%
              \let\@currentlabel=\theequation
              \global\@eqnswtrue
              \global\@eqcnt\z@
              \tabskip\@centering
              \let\\=\@eqncr
              $$%
 \halign to \displaywidth\bgroup
    \eqnumphantom\@eqnsel\hskip\@centering
    $\displaystyle \tabskip\z@ {##}$%
    &\global\@eqcnt\@ne \hskip 2\arraycolsep
         $\displaystyle\arraymode{##}$\hfil
    &\global\@eqcnt\tw@ \hskip 2\arraycolsep
         $\displaystyle\tabskip\z@{##}$\hfil
         \tabskip\@centering
    &{##}\tabskip\z@\cr}
\makeatother

\def\beq{\begin{equation}}
\def\eeq{\end{equation}}
\def\bea{\begin{eqnarray}}
\def\eea{\end{eqnarray}}

\font\teneufm=cmmib10
\font\seveneufm=cmmib7
\font\fiveeufm=cmmib5
\def\bfit#1{{\textfont1=\teneufm\scriptfont1=\seveneufm
\scriptscriptfont1=\fiveeufm
\mathchoice{\hbox{$\displaystyle#1$}}{\hbox{$\textstyle#1$}}
{\hbox{$\scriptstyle#1$}}{\hbox{$\scriptscriptstyle#1$}}}}

\def\Bf#1{\mbox{\boldmath $#1$}}

\def\balpha{{\Bf\alpha}}
\def\bbeta{{\Bf\beta}}
\def\bnu{{\Bf\nu}}
\def\bmu{{\Bf\mu}}
\def\bphi{{\Bf\phi}}
\def\bN{{\bfit N}}
\def\bM{{\bfit M}}
\def\bJ{{\bfit J}}
\def\be{{\bfit e}}
\def\bx{{\bfit x}}
\def\bt{{\bfit t}}
\def\bsN{{\bfit N}}
\def\bsM{{\bfit M}}
\def\bsx{{\bfit x}}

\def\bsalpha{{\Bf\alpha}}
\def\bsbeta{{\Bf\beta}}
\def\bsphi{{\Bf\phi}}

\def\W{{\rm W}}

\def\nn{\nonumber}
\def\mm{matrix model }
\def\mms{matrix models }

\begin{document}

\begin{titlepage}
\begin{center}
{{\it P.N.Lebedev Institute preprint} \hfill FIAN/TD-9/92\\
{\it I.E.Tamm Theory Department} \hfill ITEP-M-4/92\\
\hfill hepth@xxx/92\#\#
\begin{flushright}{July 1992}\end{flushright}
\vspace{0.1in}{\Large\bf Conformal Matrix Models as an\\
Alternative to Conventional Multi-Matrix Models}\\[.4in]
{\large  S. Kharchev, A. Marshakov, A. Mironov}\\
\bigskip {\it  P.N.Lebedev Physical
Institute \\ Leninsky prospect, 53, Moscow, 117 924, Russia},
\footnote{E-mail address: tdparticle@glas.apc.org   \&
mironov@sci.fian.msk.su}\\ \smallskip
\bigskip {\large A. Morozov}\\
 \bigskip {\it Institute of Theoretical and Experimental
Physics,  \\
 Bol.Cheremushkinskaya st., 25, Moscow, 117 259, Russia},
 \footnote{E-mail address: morozov@itep.msk.su}\\ \smallskip
\bigskip {\large S. Pakuliak}\\
 \bigskip {\it Research Institute for Mathematical Sciences\\
Kyoto University, Kyoto-606, Japan},
\footnote{E-mail address: pakuliak@kurims.kyoto-u.ac.jp}
\footnote{On leave of absence from Institute for Theoretical Physics, Kiev,
Ukraine
}} \end{center}
\bigskip \bigskip
\end{titlepage}

\newpage
\begin{abstract}
We introduce {\it conformal multi-matrix models} (CMM) as an alternative to
conventional multi-matrix model description of two-dimensional gravity
interacting with  $c < 1$  matter. We define CMM as solutions to (discrete)
extended Virasoro constraints. We argue that the
so defined alternatives of multi-matrix models represent the same
universality classes in continuum limit, while at the discrete level they
provide explicit solutions to the
multi-component KP hierarchy and by definition satisfy the discrete
$W$-constraints. We prove that discrete CMM coincide with the $(p,q)$-series
 of 2d gravity models in a {\it well}-{\it defined} continuum limit, thus
demonstrating that
they provide a proper generalization of Hermitian one-matrix model.
\end{abstract}

\newpage
\setcounter{footnote}0

\section{Introduction}

Matrix models are used nowadays to describe non-perturbative partition
functions, which interpolate between various sets of two-dimensional conformal
models coupled to 2d gravity, and thus serve to approach the old-standing goal
of devising a universal partition function of entire string theory.
Considerable progress so far is reached in unification of all tachyon-free
string models, associated with $c<1$ minimal conformal models, i.e. the models
from $(p,q)$-series plus 2d gravity. Namely, in \cite{FKN91a}
 it has been suggested
that the non-perturbative partition function ${\cal Z}_p[T_k]$, interpolating
between all the models with given $p$,\footnote{
Partition function of a given $(p,q)$ string model (i.e. the generating
functional of all the correlation functions in the model) arises from
${\cal Z}_p[T_k]$, if all the time-variables  $T_k= 0$, except for $T_1$ and
$T_{p+q}\colon  {\cal Z}_{p,q}[T] \leftrightarrow
Z_p[T_1,0,\ldots,0,T_{p+q},0,\ldots]$.
Existing formalism does not respect explicitly
the symmetry between $p$ and $q$.}
possesses the following properties:

(i)  ${\cal Z}_p[T_k]$  is a $\tau $-function of $p$-reduced KP hierarchy,
which is completely independent of all  $T_k$ with  $k=0$ {\it mod $p$};

(ii)  ${\cal Z}_p[T_k]$  satisfies an infinite set of differential equations of
the form
\beq\label{1}
{\cal W}^{(a)}_i{\cal Z}_p = 0,\ \  a=2,\ldots,p;\ \  i\geq 1-a,
\eeq
where  ${\cal W}^{(a)}_i$ are (harmonics of) the generators of Zamolodchikov's
$W_p$-algebra \cite{Zam85,FL88}, (see sect.4 below for explicit expressions
of
${\cal W}^{(a)}_i$ in terms of  $T$  and  $\partial /\partial T$). These two
statements (and actually the second one alone) are enough to define
${\cal Z}_p[T]$
completely. Further interpolation between different values of $p$ is described
in terms of Generalized Kontsevich Model (\cite{KMMMZ91a,KMMMZ91b} and
references therein).

Though the above formulated suggestion of \cite{FKN91a} can be just used as a
reasonable {\it definition} of non-perturbative partition functions (and these
are the corollaries of this definition which should be studied to proceed
further with the string program), there are still many interesting but
unresolved problems, which concern the original {\it motivation} of that
suggestion. The most direct motivation is usually supposed to come from the
study of the multi-scaling continuum limit of discrete multi-matrix models (the
number of matrices being $p-1)$. While being completely understood
in the simplest case of $p=2$ (one-matrix model) \cite{MMMM91},
such continuum limit
was never honestly studied for any $p>2$. This problem attracted certain
attention, but remains unresolved because of considerable technical
difficulties.

In the present paper we are going to argue that there is a way around all these
difficulties. Moreover, we continue to advocate the central idea of
\cite{GMMMO91}
that one has to look for the origin of all essential features of continuum
double-scaling limit of \hbox{(multi-)}
matrix models in their discrete counterparts.
{}From this point of view one accepts that the conventional multi-matrix
models,
as defined in \cite{CGM90,GM90c,BDKS90}
are not the simplest possible discrete representatives
of the relevant classes of universality. This is more or less clear already
from the fact that these do not satisfy any simple Ward identities at the
discrete level (or at least, the structure of these identities --- so called
$\tilde W$-constraints \cite{MMM92b} ---
is rather different from the $W$-constraints
(\ref{1})). Moreover, the advantage of being solved by orthogonal polynomial
technique is not enough to present them in the form which respects the whole
integrable structure.

Below we shall confirm the suggestion of \cite{MMM91}
that much better representatives
of the {\it same} universality classes are provided by somewhat different
``multi-matrix" models or just solutions to integrable hierarchies and string
equations which can be written in the multiple-integral form essentially
different from conventional multiple integrals of multi-matrix models (by
``interaction" of the nearest neighbours as well as by specific choice of
integration contours). We shall call them {\it conformal multi}-{\it matrix
models} (CMM), because of their relation to the formalism of conformal field
theory. These models are by definition constructed to satisfy  $W$-constraints
at discrete level. Moreover, some especially interesting solution to these
discrete $W$-constraints immediately at the discrete level possess a rich
{\it integrable structure} which is in some sense less rough than the
integrable structure found in the case of ordinary multi-matrix models
\cite{GMMMO91,AS91}.
More concretely, the partition function of the standard multi-matrix
models is a $\tau $-function of Toda lattice hierarchy in the first and the
last times, the other times describing only the parameterization of the point
of the Grassmannian. Thus, this case corresponds to extremely non-economic
usage of the whole variety of parameters (times) in the problem. The way to
rule out this drawback is to involve all these parameters as new times of a
more rich hierarchy, namely, some multi-component hierarchy.
It turns out that our CMM are related to the multi-component KP hierarchy of
the $SL(p)$ type (or generalized AKNS type \cite{UT84,AKNS74}).
Thus, we can claim that the proper viewpoint on the one-matrix model is to
consider it as a reduced 2-component KP hierarchy.

In what follows we shall discuss some generic (especially integrability)
properties of CMM in more details than it was done in \cite{MMM91}, and give a
detailed description of the relevant continuum limit (analogous to the
presentation of \cite{MMMM91}
for $p=2$, i.e. including the definition of Kazakov
variables  $t_k \rightarrow  T_k$, reduction, rescaling of partition function
and the transformation of discrete $W$-constraints into the continuum ones
(\ref{1}),
see also \cite{MMP92}). As a result, we argue that the continuum limit
\beq\label{2}
\lim_{N\to\infty}
\left\lbrace Z^{\rm CMM,red}_{p,N}[t_k]\right\rbrace ^{1/p} =
{\cal Z}_p[T_k]
\eeq
is exactly a solution to eqs.(\ref{1}), i.e. the relevant partition function
of the $c<1$ string theory.

The plan of the paper is as follows. In the sect.2 we remind (and make more
explicit) the construction of CMM from ref.\cite{MMM91} and discuss the form of
(Zamolodchikov's) $W$-constraints in these discrete models. In the sect.3
integrability properties of $Z^{\rm CMM}_p[t]$ are discussed. The sect.4 is
describes the $p$-reduction of CMM, the analogue of Kazakov variables and the
transformation of discrete into continuum $W$-constraints mainly along lines of
the ref.\cite{MMP92}. Concluding remarks are given in the sect.5.

\section{Conformal multi-matrix models}

In this section we shall remind the ideas of \cite{MMM91}
 and give the definition of
CMM. First, we show that the simplest example of discrete Hermitian 1-matrix
model can be easily reformulated in these terms.

Indeed, Hermitian one-matrix model $(p=2)$ can be defined as a solution to
discrete Virasoro constraints:
$$
L_nZ_{2,N}[t] = 0,\ \  \   n \geq  -1
$$
\beq\label{3}
L_n \equiv \sum ^\infty _{k=0}kt_k\partial /\partial t_{k+n} +
\sum _{a+b=n}\partial ^2/\partial t_a\partial t_b
\eeq
$$
\partial Z_{2,N}/\partial t_0 = -NZ_{2,N}
$$
The Virasoro generators (\ref{3})
 have the well-known form of the Virasoro operators in
the theory of one free scalar field. If we look for such solution in terms of
holomorphic components of the scalar field
\bea\label{4}
\phi (z) &=  \hat q + \hat p \log z  + \sum _{k\neq 0} {J_{-k}\over k}
z^{-k}\nn\\
\  [J_n,J_m] &= n\delta _{n+m,0},  \ \ \     [\hat q,\hat p] = 1
\eea
the procedure is as follows. Define vacuum states
\bea\label{5}
J_k|0\rangle  &= 0, \ \ \  \langle N|J_{-k} = 0, \ \ \    k > 0\nn\\
\hat p|0\rangle  &= 0, \ \ \   \langle N|\hat p = N\langle N|
\eea
the stress-tensor\footnote
{For the sake of brevity, we omit the sign of normal ordering in the
evident places, say, in the expression for $T$ and $W$ in terms of
free fields.}
\beq\label{6}
T(z) = {1\over 2}[\partial \phi (z)]^2 = \sum    T_nz^{-n-2},\quad
T_n = {1\over 2}\sum _{k>0}J_{-k}J_{k+n} +
{1\over 2}\sum _{{a+b=n}\atop{a,b\geq 0}}J_aJ_b,
\eeq
\beq\label{6a}
 T_n|0\rangle  = 0,  \ \ \    n \geq  -1
\eeq
and the Hamiltonian
\bea\label{7}
H(t) &= {1\over \sqrt{2}} \sum _{k>0}t_kJ_k =
\oint_{C_0}V(z)j(z)\nn\\
V(z) &= \sum _{k>0}t_kz^k, \ \  \   j(z) = {1\over \sqrt{2}}\partial \phi (z).
\eea
Now one can easily construct a ``conformal field theory" solution to
(\ref{3}) in two steps. First,
\beq\label{8}
L_n\langle N|e^{H(t)}\ldots = \langle N|e^{H(t)}T_n\ldots
\eeq
can be checked explicitly. As an immediate consequence, any correlator of the
form
\beq\label{9}
\langle N|e^{H(t)}G|0\rangle
\eeq
($N$  counts the number of zero modes of  $G$) gives a solution to (\ref{3})
provided
\beq\label{10}
[T_n,G] = 0, \ \ \  n \geq  -1
\eeq
Second, the conformal solution to (\ref{10})
(and therefore to (\ref{3})) comes from the
properties of 2d conformal algebra. Indeed, any solution to
\beq\label{11}
[T(z),G] = 0
\eeq
is a solution to (\ref{10}),
and it is well-known that the solution to (\ref{11}) is a
function of {\it screening charges}
\beq\label{12}
Q_\pm  = \oint J_\pm  = \oint
e^{\pm \sqrt{2}\phi }.
\eeq
With a selection rule on zero mode it gives
\beq\label{13}
G = \exp \ Q_+ \rightarrow  {1\over N!}Q^N_+
\eeq
(Of course, the general case must be  $G \sim  Q^{N+M}_+Q^M_-$ but the special
prescription for integration contours, proposed in \cite{MMM91},
implies that the
dependence of $M$ can be irrelevant and one can just put  $M = 0.)$
In this case the solution
\beq\label{14}
Z_{2,N}[t] = \langle N|e^{H(t)}\exp Q_+|0\rangle
\eeq
after computation of the free theory correlator, analytic continuation of the
integration contour (see more detailed discussion of this point below) gives
well-known result
\bea\label{15}
Z_{2,N} &= (N!)^{-1}\int   \prod ^N_{i=1}dz_i \exp  \left( - \sum
t_kz^k_i\right)  \Delta ^2_N(z) =\nn\\
&= (N!{\rm Vol}\ U(N))^{-1}\int   DM\ \exp \left( - \sum    t_kM^k\right)
\\
\Delta _N &= \prod ^N_{i<j}(z_i - z_j)\nn
\eea
in the form of multiple integral over spectral parameter or integration over
Hermitian matrices.

In the case of  $p=2$  (Virasoro) constraints this is just a useful
reformulation of the Hermitian 1-matrix model. However, in what follows we are
going to use this point of view as a constructive one. Indeed, instead of
considering a special direct multi-matrix generalization of (\ref{15})
\cite{CGM90,GM90c,BDKS90} one
can use powerful tools of conformal theories, where it is well known how to
generalize almost all the steps of above construction: first, instead of
looking for a solution to Virasoro constraints one can impose extended
Virasoro or  $W$-constraints on the partition function. In such case one would
get Hamiltonians in terms of {\it multi}-scalar field theory, and the second
step is generalized directly using {\it screening charges} for  $W$-algebras.
The general scheme looks as follows

(i)  Consider Hamiltonian as a linear combination of the Cartan currents of a
level one Kac-Moody algebra  ${\cal G}$
\beq\label{16}
H(t^{(1)},\ldots,t^{({\rm rank}\ {\cal G})}) =
\sum _{\lambda ,k>0}t^{(\lambda )}_k\bmu _\lambda \bJ_k,
\eeq
where $\{\bmu_i\}$ are basis vectors in Cartan hyperplane, which,
say for $SL(p)$ case are chosen to satisfy
$$
\bmu_i\cdot \bmu_j=\delta_{ij}-{1\over p},  \ \ \ \sum_{j=1}^p
\bmu_j=0.
$$

(ii)  The action of differential operators  ${\cal W}^{(a)}_i$ with respect to
times  $\{t^{(\lambda )}_k\}$ can be now defined from the relation
\beq\label{17}
W^{(a)}_i\langle \bN|e^{H(\{t\})}\ldots =
\langle \bN|e^{H(\{t\})}{\rm W}^{(a)}_i\ldots\ , \ \ \   a=2,\ldots,p;  \ \ \
i\geq 1-a,
\eeq
where
\bea\label{18}
\W^{(a)}_i &= \oint z^{a+i-1}\W^{(a)}(z)\nn\\
\W^{(a)}(z) &= \sum  _\lambda  [\bmu _\lambda \partial \bphi (z)]^a + \ldots
\eea
are  spin-$a$ W-generators of  $\W_p$-algebra written in terms of
rank$\;{\cal G}$-component scalar fields \cite{FL88}.

(iii)  The conformal solution to (\ref{1}) arises in the form
\beq\label{19}
Z^{\rm CMM}_{p,\bsN}[\{t\}] = \langle \bN|e^{H(\{t\})}G\{Q
^{(\alpha)} \}|0\rangle
\eeq
where  $G$  is an exponential function of screenings of level one Kac-Moody
algebra
\beq\label{20}
Q^{(\alpha)}  = \oint J^{(\alpha)}  = \oint e^{\bsalpha \bsphi }
\eeq
$\{\balpha \}$ being roots of finite-dimensional simply laced
Lie algebra ${\cal G}$. (For the case of non-simply
laced case see \cite{FF92}. Below ${\cal G}=SL(p)$ if not stated otherwise.)
The correlator (\ref{19}) is still a free-field correlator
and the computation gives it again in a multiple integral form
\bea\label{21}
Z^{\rm CMM}_{p,\bsN}[\{t\}] &\sim  \int   \prod  _\alpha
\left[ \prod ^{N_\alpha }_{i=1}dz^{(\alpha )}_i \exp \left( -
\sum _{\lambda ,k>0}t^{(\lambda )}_k(\bmu _\lambda \balpha )(z^{(\alpha )}_i)^k
\right) \right] \times \nn\\
&\times \prod _{(\alpha ,\beta )}\prod ^{N_\alpha }_{i=1}
\prod ^{N_\beta }_{j=1}(z^{(\alpha )}_i- z^{(\beta )}_j)^{\bsalpha \bsbeta }
\eea
The expression (\ref{21})
is what we shall study in this paper: namely the solution to
discrete  $W$-constraints (\ref{1})
which can be written as multiple integral over
spectral parameters  $\{z^{(\alpha )}_i\}$ (this integral is sometimes called
``eigenvalue model"). The difference with the one-matrix case (\ref{15})
 is that the
expressions (\ref{21}) have rather complicated representation in terms of
multi-matrix integrals. Namely, the only non-trivial (Van-der-Monde) factor can
be rewritten in the (invariant) matrix form:
\bea\label{21a}
\prod ^{N_\alpha }_{i=1} \prod ^{N_\beta }_{j=1}(z^{(\alpha )}_i-
z^{(\beta )}_j)^{\bsalpha \bbeta } = \left[ \det \{M^{(\alpha )}\otimes I -
I\otimes M^{(\beta )}\}\right] ^{\bsalpha \bsbeta },
\eea
where $I$ is the unit matrix. Still this is a model with a chain of matrices
and with closest neibour interactions only (in the case of $SL(p)$).

The purpose of this paper is to show that CMM, defined by (\ref{21}) as a
solution to the $W$-constraints has indeed a very rich integrable structure
and possesses a natural continuum limit. To pay for these advantages one
should accept a slightly less elegant matrix integral with the entries like
 (\ref{21a}).

The first non-trivial example (which we use as a demonstration making all our
statements in the paper more clear) is the $p=3$ associated with
Zamolodchikov's  $W_3$-algebra and serves as alternative to 2-matrix model.
In this particular case one obtains
\beq\label{22}
H(t,\bar t) = {1\over \sqrt{2}} \sum _{k\geq 0}(t_kJ_k + \bar t_k\bar J_k)
\eeq
\bea\label{23}
W^{(2)}_n = &L_n = \sum ^\infty _{k=0}(kt_k\partial /\partial t_{k+n} +
k\bar t_k\partial /\partial \bar t_{k+n}) +\nn\\
&+ \sum _{a+b=n}(\partial ^2/\partial t_a\partial t_b+
\partial ^2/\partial \bar t_a\partial \bar t_b)
\eea
\bea\label{24}
W^{(3)}_n = &\sum _{k,l>0}(kt_klt_l\partial /\partial t_{k+n+l} -
k\bar t_kl\bar t_l\partial /\partial t_{k+n+l}
-2kt_kl\bar t_l\partial /\partial \bar t_{k+n+l})+\nn\\
&+ 2\sum  _{k>0}\left[ \sum _{a+b=n+k}(kt_k\partial ^2/\partial t_a
\partial t_b - kt_k\partial ^2/\partial \bar t_a\partial \bar t_b -
2k\bar t_k\partial ^2/\partial t_a\partial \bar t_b)\right]  +\nn\\
&+ {4\over 3}\sum _{a+b+c=n}(\partial ^3/\partial t_a\partial t_b\partial t_c-
\partial ^3/\partial t_a\partial \bar t_b\partial \bar t_c),
\eea
where times  $t_k$ and  $\bar t_k$ correspond to the two orthogonal directions
in  $SL(3)$  Cartan plane. (We use the standard specification of the Cartan
basis:  $\be = \balpha _1/\sqrt{2}$, $\bar\be =
\sqrt{3}\bnu _2/\sqrt{2}$.
  This basis is convenient for discussing integrability
properties of the model in the sect.3, for the continuum limit we will use
another basis in Cartan plane connected with  $t \pm  \bar t$.) In this case
one has six screening charges  $Q^{(\pm \alpha _i)}$ $(i = 1,2,3)$
which commute with
\beq\label{25}
\W^{(2)}(z) = T(z) = {1\over 2}[\partial \bphi (z)]^2
\eeq
and
\beq\label{26}
\W^{(3)}(z) = \sum ^3_{\lambda =1}(\bmu _\lambda \partial \bphi (z))^3,
\eeq
where  $\bmu _\lambda $ are vectors of one of the fundamental representations
({\bf 3} or $\bar {\bf 3})$ of  $SL(3)$.

The particular form of integral representation (\ref{21})
depends on particular screening insertions to the correlator (\ref{19}).
Following ref.\cite{MMM91} we will concentrate on the solutions which have
no denominators. One of the reasons of
such choice is that these solutions possess the most simple integrable
structure of the forced hierarchy type \cite{KMMOZ91,KMMM92a}
(see sect.3), though the other ones can still be analyzed in the same manner.

One of the possible ways to avoid the problem with other solutions is the
following prescription with the integration contours. First, as in any
conformal theory, one has to consider closed contours in the definition of
screening charges (\ref{12}), (\ref{20}).
Then, the denominators can be easily integrated
out using the ordinary residue technique. However, in the final result, it is
necessary to continue contours to their specific locations defining any
concrete model. Say, in the simplest example of the Hermitian
one-matrix model
the integration runs along real line. This is not too simple procedure, because
only after this the solution starts to be non-trivial. (In a sense, with
properly defined contour  $Q^{(i)}|0\rangle  = {\displaystyle\int}
  _{\bf R}J^{(i)}|0\rangle  \neq
|0\rangle $,  what would be the case for closed contour). Thus, the
prescription to get rid of denominators implies that one first integrates over
some part of the spectral variables and then uses an analytic continuation.
However, even besides this prescription the integrable structure of generic
solutions (\ref{21})
can be analyzed in the free-fermion formalism and we shall return to this
problem in the sect.3.

The simplest solutions which have no denominators correspond to specific
correlators
\beq\label{27}
Z^{\rm CMM}_{p,\bsN}[\{t\}] = \langle \bN|e^{H(\{t\})}\prod_i
\exp Q_{\alpha _i}|0\rangle
\eeq
when we take  $\alpha _i$ to be ``neighbour" (not simple!) roots:
$(\balpha _i\balpha _j) =
1$. In the case of  $SL(3)$  this corresponds, say, to  insertions of only
  $Q_{\alpha _1}$ and  $Q_{\alpha _2}$
\bea\label{28}
&Z^{\rm CMM}_{3;M,N}[t,\bar t] \equiv  Z_{M,N}[t,\bar t] = {1\over N!M!}
\langle N,M|e^{H(t,\bar t)}(Q^{(\alpha _1)})^N(Q^{\alpha _2})^M|0\rangle  =
\nn\\
&= {1\over N!M!}\int   \prod    dx_idy_i \exp \left( - \sum    [V(x_i) +
\bar V(y_i)]\right)  \Delta ^2_N(x) \Delta ^2_M(y) \prod _{i,j}(x_i - y_j)
\eea
This expression can be examined for its integrable properties by methods
developed in \cite{KMMOZ91} (see sect.3).

Before we proceed to a detailed investigation of the integrable structure of
CMM, let us make two remarks about possible generalizations. The first one
concerns supersymmetric matrix models. This problem was first examined in
\cite{AG91}
 where the authors looked for a solution to the system of equations
 $L_nZ = 0$  and  $G_mZ = 0$,  with generators $\{L_n,G_m\}$  forming  the
$N = 1$  superconformal algebra. The solution was found
in the form of a multiple integral over even and odd (Grassmannian) parameters,
in the form similar to that considered above. It is necessary to point out that
in our language this is nothing but trivial generalization of the
one-field case which has to be substituted by scalar superfield.
Then the insertion of screenings of  $N = 1$  superconformal algebra
immediately leads to the result by \cite{AG91}.
{}From this point of view the real problem with supersymmetric generalization
can arise only in the  $N = 2$ case because of the lack of appropriate
screening operators.

The second possible generalization is to the case of ``deformed'' Virasoro$-W$
constraints via Feigin-Fuchs-Dotsenko-Fateev procedure. In such case formulas
(\ref{14}) and (\ref{19})
 are still valid, one has only change the definitions (\ref{6}), (\ref{18}) to
\beq\label{29}
T = {1\over 2}[\partial \phi (z)]^2 + \alpha _0\partial ^2\phi
\eeq
(and corresponding formulas for higher W-algebras with  $c < {\rm rank}
{\cal G})$
with corresponding deformations of generators (\ref{1}) written in terms of
differential operators. Deformed generators will now commute with new screening
charges
\beq\label{30}
Q_\pm  = \oint e^{\alpha _\pm \phi }, \ \ \
\alpha _\pm  = \alpha _0 \pm  (\alpha _0 + 2)^{1/2}
\eeq
in the case of deformed  $p=2$  theory (minimal models). The only problem
which might arise here is connected with the integrability of such system ---
expressions (\ref{30})
have no natural representation via free $fermions$ for generic
$\alpha _0$. To avoid this problem one should consider
a $j$-differential system \cite{GO89}. After the Miwa transform \cite{Miw82}
of the time-variables
\beq\label{31}
t_k = {1\over k}\sum_j a_j\xi ^{-k}_j
\eeq
our correlators acquire the form of
\beq\label{32}
\langle e^{H(\{t\})}Q^N_+Q^M_-\rangle  = {\langle
\prod _i \Phi _i(\xi _i)Q^N_+Q^M_-\rangle \over
\prod _{i<j}(\xi_i-\xi_j)^{a_ia_j}}
\eeq
with
\beq\label{33}
\Phi _i(z) = \exp [a_i\phi (z)], \ \ \  \Delta _i = a^2_i/2 - a_i\alpha _0
\eeq
One can choose  $\{a_j\}$  in such a way that  $\{\Phi _j(z)\}$  become
primaries of minimal models, in such case the correlators (\ref{32})
will satisfy
certain differential equations
 following from the null-vector conditions (in the particular case of
 2-level degeneration they will be identical with the Virasoro constraints
  in Miwa variables).
 This can be also easily generalized to the $W$-case of several scalar
 fields and several sets of times. In this case one should
 introduce in (\ref{31}) {\em vectors} $\bt_k$ and $\balpha_j$.
 This leads to independent differential equations which again correspond
 to the independent null-vector conditions.
Such construction trivially reveals the origin of the correspondence between
Virasoro constrained $\tau $-functions and correlators in the minimal models.

Unfortunately the construction is valid in its present form only for
{\it discrete} models (solution to discrete constraints) and is rather
difficult to
generalize to continuum models. The reason is that in the latter case we
have to impose the twisted boundary conditions on scalar fields and thus
(i) there is nothing like selection rule (\ref{13}) which in the
discrete case is provided by a zero mode and
(ii) eq.(\ref{6a}) is no longer satisfied. We are going to return to this
problem in a separate publication (see also \cite{Ger91}).

\section{Integrability of conformal multi-matrix models}

In this section we are going to present a detailed investigation of integrable
structure of CMM, based on the application of the general technique of
\cite{KMMOZ91}.
In the subsect.3.1 we will derive a determinant representation for CMM
partition function which is an indication to the fact that we deal with an
integrable system and the partition function is a certain $\tau $-function. As
a direct result of the determinant formula we obtain the Hirota bilinear
relation satisfied by partition function of CMM, which is a generalization of
that one for the Hermitian one-matrix model.
In the subsect.3.2 we are going to
apply the fermionic operator formalism for CMM which brings us to a conclusion
that the partition function of CMM is a $\tau $-function of a multi-component
Kadomtsev-Petviashvili hierarchy which obeys the constraint
\beq\label{34}
\sum ^p_{k=1}\partial /\partial t^{(k)}_n\tau ^{(p)}(\{t\}) = 0,  \ \ \
   n =  1,2,\ldots\  ,
\eeq
where  $p$  is the number of components. This $\tau $-function satisfies
$W$-constraints (\ref{1})
and is a discrete counterpart of the analogous statement in
the continuum limit found in \cite{FKN91a}.

\subsection{Determinant representation}

Let us first remind briefly the results of \cite{KMMOZ91}
 for the simplest example of
$p=2$. The partition function of $p=2$ case (\ref{15})
 can be written in the form
\beq\label{35}
Z_{2,N}(t) = \det _{N\times N}[\partial ^{i+j-2}C(t)] = \tau _N(t)
\eeq
with
\beq\label{36}
\partial _{t_n}C(t) = \partial ^n_{t_1}C(t)
\eeq
which implies that it is a $\tau $-function of the Toda chain hierarchy (or
a $\tau $-function of AKNS-reduction of 2-component KP hierarchy \cite{New85};
 see ref.
\cite{KMMOZ91} and sect.3.2 below for details). Eq.(\ref{36})
means that  $C(t)$  just has an integral representation
\beq\label{37}
C(t) = \int   d\mu (z) \exp  \sum    t_kz^k,
\eeq
where $d\mu $ is some measure; these are the Virasoro constraints which fix
 the concrete measure and the contour of integration in (\ref{37}).
The determinant form (\ref{35}) is an explicit
manifestation of the fact that the partition function does satisfy the Hirota
bilinear relations, the simplest one of which in this particular case takes the
form
\beq\label{38}
\tau _N(t) {\partial ^2\over \partial t^2_1}\tau _N(t) -
\left( {\partial \tau _N(t)\over \partial t_1}\right) ^2 =
\tau _{N+1}(t)\tau _{N-1}(t)
\eeq
The fact that in eq.(\ref{35}) we are dealing with a (forced) Toda chain
reduction of generic Toda lattice (two-component
KP hierarchy) is reflected in the specific feature of the matrix
in (\ref{35}): $C_{ij} = C_{i+j}$ \cite{UT84,KMMOZ91,KMMM92a}.

Now we are going to generalize (\ref{35}) and (\ref{38}).
In the case of the $p=3$ model (\ref{28})  one has to introduce two functions
instead of (\ref{37}):
\beq\label{39}
C(t) = \int   dz\ \exp  [-V(z)],  \ \ \    \bar C(\bar t) = \int   dz\ \exp
[-\bar V(z)]
\eeq
where
$$
V(z) = \sum _{k>0}t_kz^k, \ \ \     \bar V(z) = \sum _{k>0}\bar t_kz^k
$$
and
\beq\label{40}
\partial _{t_n}C(t) = \partial ^n_{t_1}C(t), \ \ \
\partial _{\bar t_n}
\bar C(\bar t) = \partial ^n_{\bar t_1}\bar C(\bar t)
\eeq
The determinant representation has now the form  $(\partial  \equiv
\partial /\partial t_1$,  $\bar \partial  \equiv  \partial /\partial \bar t_1)$
$$
Z_{N,M}(t,\bar t) =
$$
$$
= \det \left[{\small
\begin{array}{cccccccc}
C&\partial C&\ldots&\partial^{N-1}C&
\bar C&\bar\partial \bar C&\ldots&
\bar\partial^{M-1}\bar C\\
\partial C&\partial^{2}C&\ldots&\partial^{N}C&
\bar\partial \bar C&\bar\partial^{2}\bar C&\ldots&
\bar\partial^{M}\bar C\\
&\phantom{a}&&&&&&\\
\partial^{N+M-1}C&\partial^{N+M}C&\ldots&\partial^{2N+M-2}C&
\bar\partial^{N+M-1}\bar C&\bar\partial^{N+M}\bar C&\ldots&
\bar\partial^{2N+M-2}\bar C
\end{array}}\right]=
$$
\beq\label{41}
\equiv  \tau _{N,M}(t,\bar t)
\eeq
which is exactly the double-Wronskian representation of a
$\tau $-function \cite{Hir}.

We can prove the determinant representation (\ref{41})
as follows. Instead of eq.(\ref{28})
let us consider slightly more general expression by introducing the external
sources $\{\beta _i\}$ and $\{\bar \beta _i\}$:
\bea\label{42}
&Z^{\rm CMM}_{3;M,N}[t,\bar t\mid\beta ,\bar \beta ] =
{1\over N!M!}\nn\\
&\times \int   \prod    dx_idy_i \exp \left( - \sum    [V(x_i) -
\beta _ix_i+ \bar V(y_i) - \bar \beta _iy_i]\right)  \Delta _N(x) \Delta _M(y)
\Delta (x,y),
\eea
where  $\Delta (x,y) \equiv  \Delta _N(x) \Delta _M(y) \prod _{i,j}(x_i - y_j)$
is  $(N+M)\times (N+M)$ Van-der-Monde determinant. Using the derivatives with
respect to $\beta _i$ and $\bar \beta _i$ one can get rid of $\Delta (x,y)$ in
the integrand thus obtaining
\beq\label{43}
Z^{\rm CMM}_{3;M,N}[t,\bar t\mid \beta ,\bar \beta ] =
\Delta (\partial _\beta ,\partial_{\bar \beta }
) F(t,\beta )\bar F(\bar t,\bar \beta ),
\eeq
where $F$ and $\bar F$ have the specific form of Kontsevich-like integrals :
$$
F(t,\beta ) = \int   \prod    dx_i \exp \left( - \sum    [V(x_i) -
\beta _ix_i]\right)  \Delta _N(x)\   ,
$$
$$
\bar F(\bar t,\bar \beta ) = \int   \prod    dx_i \exp \left( - \sum
[\bar V(x_i) - \bar \beta _ix_i]\right)  \Delta _N(x)\  .
$$
Using the trick with the differentiation over $\beta $ and $\bar \beta $ again
one can represent these expressions in the determinant form (compare with
\cite{KMMMZ91a,KMMMZ91b}):
\beq\label{44}
F(t,\beta ) = \det [ \partial ^{i-1}_{\beta _j}C(t,\beta _j)] = \det [
\partial ^{i-1}_{t_1}C(t,\beta _j)],\ \ \  i,j = 1,\ldots,N,
\eeq
\beq\label{45}
\bar F(\bar t,\bar \beta ) = \det [ \partial ^{i-1}_{\bar\beta_i}
\bar C(\bar t,\bar \beta _j)] = \det [ \partial ^{i-1}_{\bar t_1}
\bar C(\bar t,\bar \beta _j)],  \ \ \ i,j = 1,\ldots,M,
\eeq
where
$$
C(t,\beta ) \equiv  \int   dz\ \exp  [- V(z) + \beta z],  \ \ \
\bar C(\bar t,\bar \beta ) \equiv  \int   dz\ \exp  [- \bar V(z) +
\bar \beta z].
$$

Substitution of eqs.(\ref{44}) and (\ref{45})
into eq.(\ref{43}) leads to the expression which
contains the sum of $M!N!$ terms analogous to (\ref{41})
 (each term corresponds to a
particular permutation of $\{\beta _i\}$ and $\{\bar \beta _i\})$. Finally, in
the limit of $\beta _i = \bar \beta _i = 0$ it reproduces the eq.(\ref{41}).

Now the generalization of (\ref{41}) for $p \geq  3$ is quite obvious:
$\tau $-function has ``$(p-1)$-tuple" Wronskian form for
$(\sum \ N_i)\times (\sum \ N_i)$ matrix (multiplied by the factor $\prod
N_i!)$ with corresponding  $C_i(t_i) = {\displaystyle\int}
   \exp [ - V_i(z)]dz$ $(i = 1,
\ldots,p-1)$.

{}From representation (\ref{41}) it is easy to derive that
\beq\label{46}
{\partial ^2\over \partial t_1\partial \bar t_1}\log \tau _{N,M}(t,\bar t) =
{\tau _{N+1,M-1}(t,\bar t)
\tau _{N-1,M+1}(t,\bar t)\over \tau ^2_{N,M}(t,\bar t)}
\eeq

This expression can be also easily extended to the $(p-1)$-matrix case:
\beq\label{47}
{\partial ^2\over \partial t^{(i)}_1\partial t^{(j)}_1}
\log \ \tau _{\{N_k\}}(t,\bar t) =
{\tau _{\{..,N_i+1,..,N_j-1,..\}}(t,\bar t)
\tau _{\{..,N_i-1,..,N_j+1,..\}}(t,\bar t)\over \tau ^2_{\{N_k\}}(t,\bar t)}
\eeq

This equations of motion can be immediately derived from the corresponding
determinant representation like the case of Wronskian solutions of KP- or Toda
lattice hierarchies (see, for example, \cite{Hir})
 and are the first Hirota bilinear
equations generalizing (\ref{38}).

\subsection{Fermionic representation}

Now we shall proceed to the representation of the solutions to CMM in terms of
free fermion correlation functions. Such a representation (invented for
integrable hierarchies by Kyoto school \cite{DJKM83})
allows one to establish some
of the properties of the system under consideration in an elegant and
``physical" way.

Again, first we are going to show that the solution to Hermitian one-matrix
model is nothing but AKNS-reduction of 2-component KP hierarchy
\cite{GMMMO91,KMMOZ91}. Indeed, the $\tau $-function of 2-component KP
hierarchy is by definition the correlator
\beq\label{3.2.1}
\tau ^{(2)}_{N,M}(x,y) = \langle N,M|e^{H(x,y)}G|N+M,0\rangle
\eeq
where
\beq\label{3.2.2}
H(x,y) = \sum _{k>0}(x_kJ^{(1)}_k + y_kJ^{(2)}_k)
\eeq
\beq\label{3.2.3}
J^{(i)}(z) = \sum    J^{(i)}_kz^{-k-1} = {:}\psi
^{(i)}(z) \psi ^{(i)\ast }(z){:}
\eeq
\beq\label{3.2.4}
\psi ^{(i)}(z) \psi ^{(j)\ast }(z') = {\delta _{ij}\over z - z'} + \ldots\ \ .
\eeq
Now we are going to demonstrate that (\ref{14}) is equivalent to (\ref{3.2.1})
 for certain
$G$  for which (\ref{3.2.1}) depends only on the differences  $x_k-y_k$. To do
this we have to make use of the free-fermion representation of  $SL(2)_{k=1}$
Kac-Moody algebra:
\bea\label{3.2.5}
&J_0 = {1\over 2}(\psi ^{(1)}\psi ^{(1)\ast } - \psi ^{(2)}\psi ^{(2)\ast }) =
{1\over 2}(J^{(1)} - J^{(2)})\nn\\
&J_+ = \psi ^{(2)}\psi ^{(1)\ast } \quad J_- = \psi ^{(1)}\psi ^{(2)\ast }
\eea
Now let us take $G$ to be the following exponent of a quadratic form
\beq\label{3.2.6}
G \equiv  {:}\exp  \left( \int   \psi ^{(2)}\psi ^{(1)\ast }\right){:}
\eeq
The only term which contributes into the correlator (\ref{3.2.1})
due to the charge
conservation rule is:
\beq\label{3.2.7}
G_{N,M} \equiv  G_{N,-N}\delta_{M,-N} = {1\over N!}\ {:}\left( \int
\psi ^{(2)}\psi ^{(1)\ast }\right) ^N{:}~\delta_{M,-N}
\eeq
Now we bosonize the fermions
\bea\label{3.2.8}
\psi ^{(i)\ast } = e^{\phi _i},  \ \ \  \psi ^{(i)} = e^{- \phi _i}
\nn\\
J^{(1)} = \partial \phi _1, \ \ \     J^{(2)} = \partial \phi _2
\eea
and compute the correlator
$$
\tau ^{(2)}_N(x,y) \equiv  \tau ^{(2)}_{N,-N}(x,y) = {1\over N!}
\langle N,-N|\exp \left( \sum _{k>0}(x_kJ^{(1)}_k +
y_kJ^{(2)}_k)\right) \left( \int
{:}\psi ^{(2)}\psi ^{(1)\ast }{:}\right) ^N|0\rangle  =
$$
$$
= {1\over N!} \langle N,-N|\exp \left(
\oint
[X(z)J^{(1)}(z) + Y(z)J^{(2)}(z)]\right) \left( \int
:\exp (\phi _1-\phi _2):\right) ^N|0\rangle
$$
Introducing the linear combinations  $\sqrt{2}\phi  = \phi _1 - \phi _2$,
$\sqrt{2}\tilde \phi  = \phi _1 + \phi _2$ we finally get
\bea\label{3.2.9}
&\tau ^{(2)}_N(x,y) = {1\over N!} \langle \exp \left( {1\over \sqrt{2}}
\oint[X(z)+Y(z)]\partial \tilde \phi (z)\right) \rangle  \times\nn\\
&\times  \langle N|\exp \left( {1\over \sqrt{2}}
\oint [X(z)-Y(z)]\partial \phi (z)\right) \left( \int
:\exp \sqrt{2}\phi :\right) ^N|0\rangle  = \tau ^{(2)}_N(x-y)
\eea
since the first correlator is in fact independent of $x$ and $y$. Thus, we
proved that the $\tau $-function (\ref{3.2.1})
indeed depends only on the {\it difference} of two sets
of times  $\{x_k-y_k\}$. So, we obtained here a particular case of the
2-component KP hierarchy (\ref{3.2.1}) and

(i)  requiring the elements of Grassmannian to be of the form (\ref{3.2.6})
we actually performed a reduction to the 1-component case\footnote{
Note that the idea to preserve both indices in (\ref{3.2.1}) leads
immediately to additional insertions either of  $\psi^{(1)\ast}$ or
$\psi ^{(2)}$ to the right vacuum $|0\rangle $, so that it is no longer
annihilated at least by  the $T_{-1}$ Virasoro generator, or in other words
this ruins the string equation. Thus only the particular reduction
(\ref{3.2.6}) seems to be
consistent with string equation. This choice of indices just corresponds to
that considered originally in \cite{DJKM81}.};

(ii)  we proved in (\ref{3.2.9}) that this is an AKNS-type reduction for
the $\tau $-function (\ref{3.2.8}) \cite{UT84,New85}.

The above simple example already contains all the basic features of at least
all the  $A_p$ cases. Indeed, the reduction (\ref{3.2.9})
is nothing but  $SL(2)$-reduction
of a generic  $GL(2)$  situation. In other words, the diagonal  $U(1)$
$GL(2)$-current  $\tilde J = {1\over 2}(J^{(1)} + J^{(2)}) =
{1\over \sqrt{2}}\partial \tilde \phi $  decouples. This is an invariant
statement which can be easily generalized to higher  $p$  cases.

In the case of  $SL(p)$  we have to deal with the $p$-component hierarchy and
instead of (\ref{3.2.1}) for generic $\tau $-function one has
\bea\label{3.2.10}
\tau ^{(p)}_N(x) &= \langle N|e^{H(x)}G|0\rangle\\
N &= \{N_1,\ldots,N_p\}, \ \ \     x = \{x^{(1)},\ldots,x^{(p)}\}\nn
\eea
and now we have  $p$  sets of fermions  $\{\psi ^{(i)\ast },\psi ^{(i)}\}$
$i=1,...,p$. The Hamiltonian is given by Cartan currents of $GL(p)$
\bea\label{3.2.11}
H(t) &= \sum ^p_{i=1} \sum _{k>0}x^{(i)}_kJ^{(i)}_k\\
J^{(i)}(z) &= \psi ^{(i)}\psi ^{(i)\ast }(z)\nn
\eea
and the element of the Grassmannian in the particular case of CMM is given by
an exponents of the other currents
\beq\label{3.2.12}
J^{(ij)} = \psi ^{(i)}\psi ^{(j)}, \ \ \  \tilde J^{(ij)} =
\psi ^{(i)}\psi ^{(j)\ast },  \ \ \  J^{(ij)\ast } =
\psi ^{(i)\ast }\psi ^{(j)\ast },  \ \ \    i \neq  j
\eeq
i.e.
\bea\label{3.2.13}
G &\equiv  \prod     \exp (Q^{(ij)})\exp (\tilde Q^{(ij)})\exp (Q^{(ij)\ast })
\\
Q^{(ij)} &=  \oint
J^{(ij)},  \  \ \   \tilde Q^{(ij)} =\oint
\tilde J^{(ij)}, \ \ \   Q^{(ij)\ast } = \oint
J^{(ij)\ast }, \ \ \       i \neq  j \nn
\eea
Since  (\ref{3.2.12}) are the $SL(p)_1$ Kac-Moody currents,(\ref{3.2.13}) play
the role of screening operators in the theory under consideration.
It deserves mentioning that they are exactly  the $SL(p)$ (not  $GL(p))$
-screenings and thus the $\tau $-function (\ref{3.2.10}) does not depend on
$\{\sum ^p_{i=1}x^{(i)}_k\}$, i.e. we obtain the constraint (\ref{34}).

In the case of  $SL(3)$  this looks as follows. The screenings are
\beq\label{3.2.14}
Q^{(\bsalpha )} =\oint
J^{(\bsalpha )},
\eeq
where  $\{\balpha \}$  is the set of the six roots of  $SL(3)$. In terms of
fermions or bosons the screening currents look like
\bea\label{3.2.15}
J^{(\bsalpha _1)} &= \psi ^{(1)\ast }\psi ^{(2)\ast } = \exp (\phi _1 + \phi _2
)
\nn\\
J^{(\bsalpha _2)} &= \psi ^{(2)\ast }\psi ^{(3)\ast } = \exp (\phi _2 + \phi _3
)
\nn\\
J^{(\bsalpha _3)} &= \psi ^{(1)}\psi ^{(3)\ast } = \exp (\phi _3 - \phi _1)
\nn\\
J^{(-\bsalpha _1)} &= \psi ^{(1)}\psi ^{(2)} = \exp (- \phi _1 - \phi _2)
\\
J^{(-\bsalpha _2)} &= \psi ^{(2)}\psi ^{(3)} = \exp (- \phi _2 - \phi _3)
\nn\\
J^{(-\bsalpha _3)} &= \psi ^{(3)}\psi ^{(1)\ast } = \exp (\phi _1 - \phi _3)
\eea
The particular $\tau $-function is now described in terms of the correlator
\beq\label{3.2.16}
\tau ^{(3)}_\bsN(\bx) = \langle \bN|e^{H(\bsx)}G|0\rangle
\eeq
with
\beq\label{3.2.17}
G \sim  \prod  _\bsalpha \exp Q^{(\bsalpha )}
\eeq
The condition of Cartan neutrality is preserved by compensation of charges
between the operator (\ref{3.2.17}) and left vacuum
  $\langle \bN|$  in (\ref{3.2.16}). It
is obvious that in such case due to the condition of Cartan neutrality of the
correlator (like in Wess-Zumino models) the mode  $\tilde J =
\partial \tilde \phi  = {\displaystyle1\over
\displaystyle \sqrt{p}}\sum ^p_{i=1}\partial \phi _i$
decouples from the correlator, and
\bea\label{3.2.18}
\tau^{(3)}_\bsN(\bx)
&= \langle \bN|e^{H(\bsx)}G|0\rangle  =\nn\\
&= \left|\langle 0|\exp \left( \sum_{k>0}
\tilde J_k \sum^3_{i=1} x^{(i)}_k \right)
|0\rangle_{\tilde\phi}
 \langle \bN|e^{H(t,\bar t)}G|0\rangle\right|_{\sum_i\phi_i=0}=\\
&= \tau^{\rm red}_N(t,\bar t),
\eea
where the first correlator in the second row is trivially equal to unity. For
the specific choice of the operator  $G$  in (\ref{3.2.18})
\beq\label{3.2.19}
G = G_{1,2} = \exp \left( \int   J^{(\bsalpha _1)}\right) \exp \left( \int
J^{(\bsalpha _2)}\right)
\eeq
we reproduce the formula (\ref{41}).

Thus, the general statement is that for  $A_p$-series we obtain an ``AKNS-type"
$SL(p)$-reduction of $p$-component KP hierarchy. Generalization to other groups
 is easy. Indeed, starting from the corresponding
$W$-algebra we automatically end up with the element of the Grassmannian with
proper group properties. Thus, we have a tool to construct hierarchies of the
matrix model type (i.e. satisfying a string equation) with a given symmetry
group. All properties of these hierarchies are automatically dictated by the
symmetry. Say, the defining property of (generalized) AKNS hierarchy is
the independence of the corresponding $p$-component KP $\tau $-function of the
sum of times \cite{UT84}. On the other hand, it really corresponds to $SL(p)$
reduction from theoretical group point of view \cite{New85}.

It is even more interesting to remark again that the
general AKNS system is defined
for arbitrary set of zero times:
$$
\tau _{\bsN,\bsM} \sim  \langle \bN|\ldots|\bM\rangle
$$
with the only restriction  $\sum N_i=\sum M_i$ due to the charge conservation
law. But the condition that $T_{-1}$ annihilates the right vacuum requires
$\bM$  to be zero (it is interesting to note that the original paper
\cite{DJKM81}, devoted to multi-component KP hierarchy, was dealing with
exactly this type of restriction). In the $SL(2)$ case this gives rise to a
one-parameter $\tau $-function $\tau _{\bsN,-\bsN}$, what allows one to
rewrite the system as a Toda (chain)-type
hierarchy, since the Toda-type system should depend only on a single zero-time
(in this case the determinant of the type (\ref{41}) can be re-ordered to have
a block form with the given symmetry properties between different blocks).
In particular, the first equation from the 2-component hierarchy which is of
the form (\ref{46}) \cite{DJM82}, in this case
transforms into eq.(\ref{38}) which is just a Toda chain equation.

As to the Toda-like representation of CMM, in the simplest $SL(2)$-case
the result is indeed equivalent to the Toda chain hierarchy
\cite{GMMMO91,KMMOZ91,KMMM92a}. In
the fermionic language this connection is established by the following
substitution in the element of the Grassmannian
\beq\label{3.2.20}
\psi ^{(1)}(z) \rightarrow  \psi (z), \ \ \
\psi ^{(2)}(z) \rightarrow  \psi ({1\over z})
\eeq
and the same for $\psi ^\ast$'s. This is a reflection of the fact that Toda
system is described by the two marked points (say, 0 and $\infty$) and
corresponds to two glued discs, so it can be also described by two different
fermions. This might lead to a general phenomenon, when any multi-component
solution to CMM is actually related to (some reduction) of a multi-component
Toda lattice.

\section{Double-scaling limit}

In this section we consider the central issue  of the connection between
\mms and $2d$ gravity theory -- continuum limit. Even in the most
investigated case of Hermitian one-\mm this is a rather sophisticated
procedure, especially if one wants to reproduce the whole continuum
integrable structure of \cite{FKN91a}. Moreover, in the case of conventional
multi-\mm this procedure is still unknown.

Below we demonstrate that the advantages of CMM make it possible to define
honestly the double-scaling limit in these theories
along the line of \cite{MMMM91} (see also \cite{MMP92}).
This means that the set of times of discrete
model undergo a Kazakov-like transformation to the
continuum times.  Discrete $W$-constraints are transformed into constraints
(\ref{1}) of continuous model.

\subsection{Results of [7] for the one \mm}

To begin with let us briefly remind the main points of \cite{MMMM91}.

It has been suggested in \cite{FKN91a} that the square root of the
partition function of the continuum limit of
one-\mm is subjected to the Virasoro constraints
\beq\label{b1}
{\cal L}_n^{\rm cont}\sqrt{Z^{\rm ds}}=0,\quad n\geq -1,
\eeq
where
\bea \label{b2}
{\cal L}^{\rm cont}_n&=\sum_{k=0}\left(k+{1\over2}\right)
T_{2k+1}{\partial\over\partial T_{2(k+n)+1}}+G
\sum_{0\leq k\leq n-1}{\partial^2\over\partial T_{2k+1}\partial
T_{2(n-k-1)+1}}+\nn\\
&+{\delta_{0,n}\over16} +{\delta_{-1,n}T_1^2\over(16G)}
\eea
are modes of the stress tensor
\beq\label{b3}
{\cal T}(z)= {1\over 2}{:}\partial\Phi^{2}(z){:} - {1\over16z^2}
=\sum{{\cal L}_n\over z^{n+2}}.
\eeq

It was shown in  \cite{MMMM91} that these equations which reflect the
$W^{(2)}$-invariance of the partition function of the continuum model can
be deduced from analogous constraints in Hermitian one-\mm by taking the
double-scaling continuum limit.
The procedure (generalized below to CMM) is as follows.

The partition function of Hermitian one-\mm  can be written in the form
\beq	\label{b4}
Z\{t_{k}\}=\int{\cal D}M\exp{\rm Tr}\sum_{k=0}t_{k}
M^{k}
\eeq
and satisfies \cite{MM90,AJM90} the discrete Virasoro constraints  (\ref{3}).
\bea  \label{b5}
&L^{\rm H}_{n}Z=0,\quad n\geq0 \nn\\
&L^{\rm H}_{n}=\sum_{k=0}kt_{k}{\partial\over\partial t_{k+n}}+
\sum_{0\leq k\leq n}{\partial ^2\over \partial t_{k}\partial t_{n-k}}.
\eea
In order to obtain the above-mentioned relation between $W$-invariance of the
discrete and continuum models one has to consider a reduction of  model
(\ref{b4}) to the pure even potential $t_{2k+1}=0$.

Let us denote by the $\tau_N^{\rm red}$ the partition function of the
reduced \mm
\beq	\label{a1}
\tau^{\rm red}_{N}\{t_{2k}\}=\int{\cal D}M\exp{\rm Tr}\sum_{k=0}t_{2k}
M^{2k}
\eeq
and consider the following change of the time variables
\beq\label{a5}
g_m=\sum_{n\geq m}{(-)^{n-m}\Gamma\left(n+{3\over2}\right)
a^{-n-{1\over2}}\over(n-m)!\Gamma\left(m+{1\over2}\right)}T_{2n+1},
\eeq
where $g_m \equiv mt_{2m}$ and this expression can be used also for the zero
discrete time $g_0 \equiv N$ that plays the role of the dimension of matrices
in the one-matrix model. Derivatives with respect to  $t_{2k}$ transform as
\beq\label{a6}
{\partial\over\partial t_{2k}}=\sum_{n=0
}^{k-1}{\Gamma\left(k+{1\over2}\right)
a^{n+{1\over2}}\over(k-n-1)!\Gamma\left(n+{3\over2}\right)}{\partial
\over\partial \tilde T_{2n+1}},
\eeq
where the auxiliary continuum times $\tilde T_{2n+1}$ are connected with
``true'' Kazakov continuum times $T_{2n+1}$ via
\beq\label{a7}
T_{2k+1}=\tilde T_{2k+1}+a{k\over k+1/2}\tilde T_{2(k-1)+1},
\eeq
and coincide with $T_{2n+1}$ in the double-scaling limit when $a\to0$.

Let us rescale the partition function of the reduced one-\mm by
exponent of quadratic form of the auxiliary times $\tilde T_{2n+1}$
\beq\label{a11}
\tilde\tau=\exp\left(-{1\over2}\sum_{m,n\geq0}A_{mn}\tilde T_{2m+1}
\tilde T_{2n+1} \right)\tau^{\rm red}_N
\eeq
with
\beq\label{a12}
A_{nm}={\Gamma\left(n+{3\over2}\right)\Gamma\left(m+{3\over2}\right)\over
2\Gamma^2\left({1\over2}\right)}
{(-)^{n+m}a^{-n-m-1}\over n!m!(n+m+1)(n+m+2)}.
\eeq
Then a direct though tedious calculation \cite{MMMM91} demonstrates that
the relation
\beq\label{a15}
{\tilde{\cal L}_n\tilde\tau\over\tilde\tau}
=a^{-n}\sum_{p=0}^{n+1}C^p_{n+1}(-1)^{n+1-p}
{L_{2p}^{\rm red}\tau^{\rm red}\over \tau^{\rm red}},
\eeq
is valid, where
\bea
L_{2n}^{\rm red} \equiv \sum_{k=0}kt_{2k}{\partial\over\partial t_{2(k+n)}}+
\sum_{0\leq k\leq n}{\partial ^2\over \partial t_{2k}\partial t_{2(n-k)}}
\eea
and
\bea\label{b6}
\tilde{\cal L}_{-1}=&\sum_{k\geq1}\left(k+{1\over2}\right)T_{2k+1}
{\partial\over\partial \tilde T_{2(k-1)+1}}+
{T_1^2\over16G},\nn\\
\tilde{\cal L}_{0}=&\sum_{k\geq0}\left(k+{1\over2}\right)T_{2k+1}
{\partial\over\partial \tilde T_{2k+1}},\nn\\
\tilde{\cal L}_n=&\sum_{k\geq0}\left(k+{1\over2}\right)T_{2k+1}
{\partial\over\partial \tilde T_{2(k+n)+1}} \nn\\
&+\sum_{0\leq k \leq n-1}{\partial\over\partial \tilde T_{2k+1}}
{\partial\over\partial \tilde T_{2(n-k-1)+1}} -{(-)^n\over16a^n},\ \ \
n\geq1.
\eea
Here $C^p_n =\frac{n!}{p!(n-p)!}$ are binomial coefficients.

These Virasoro generators differ from the Virasoro generators
(\ref{b2})  \cite{FKN91a,DVV91a}
by terms which are singular in the limit $a\longrightarrow 0$.
At the same time $L_{2p}^{\rm red}\tau^{\rm red}$ at the r.h.s. of (\ref{a15})
do
not need to vanish, since
\bea\label{15aa}
0 = L_{2p}\tau\left\vert_{t_{2k+1}=0} =
L_{2p}^{\rm red}\tau^{\rm red} +
\sum_i {\partial^2\tau\over\partial t_{2i+1}\partial t_{2(n-i-1)+1}}
\right\vert_{t_{2k+1}=0}.
\eea
It was shown in \cite{MMMM91} that these two origins of difference between
(\ref{b2}) and (\ref{b6}) actually cancel each other, provided eq.(\ref{a15})
is rewritten in terms of the square root $\sqrt{\tilde\tau}$ rather than
$\tilde\tau$ itself:
\beq\label{a16}
{{\cal L}^{\rm cont}_n\sqrt{\tilde\tau}\over\sqrt{\tilde\tau}}
=a^{-n}\sum_{p=0}^{n+1}C^p_{n+1}(-1)^{n+1-p}\left. {L_{2p}\tau\over\tau}
\right\vert_{t_{2k+1}=0}\left( 1+  O(a) \right).
\eeq
The proof of this cancelation, as given in  \cite{MMMM91}, is not too much
simple and makes use of integrable equations for $\tau$.

In our consideration of  CMM below we will use a more economical way to
define the change of the time-variables $t \longrightarrow T$
(also proposed in \cite{MMMM91}), implied by the scalar field formalism.
The Kazakov change of the time variables (\ref{a5},\ref{a6}) can be
deduced from the following prescription. Let us consider the free
 scalar field with periodic boundary conditions ((\ref{d7}) for $p=2$)
\beq\label{a8}
\partial\varphi(u)=\sum_{k\geq0}g_ku^{2k-1}+
\sum_{k\geq1}{\partial\over\partial t_{2k}}u^{-2k-1},
\eeq
and analogous scalar field with antiperiodic boundary conditions:
\beq\label{a9}
 \partial\Phi(z)=\sum_{k\geq0}\left(\left(k+{1\over2}\right)T_{2k+1}
z^{k-{1\over2}}+
{\partial\over\partial \tilde T_{2k+1}}z^{-k-{3\over2}}\right).
\eeq
Then the equation
\beq\label{a10}
{1\over\tilde\tau}\partial\Phi(z)\tilde\tau =
a {1\over\tau^{\rm red}}\partial\varphi(u)\tau^{\rm red},\quad
u^2=1+az
\eeq
generates the correct transformation rules
(\ref{a5}), (\ref{a6}) and gives rise to the expression
for $A_{nm}$ (\ref{a12}).
Taking  the square of the both sides of the identity (\ref{a10}),
\bea\label{a100}
&{1\over\tilde\tau}{\cal T}(z)\tilde\tau={1\over \tau^{\rm red}}T(u)\tau^{\rm
red},
\eea
one can obtain after simple  calculations that the
same relation (\ref{a15}) is valid.

\subsection{On the proper basis for CMM}

In this subsection we would like to discuss briefly the manifest expressions
for constraint algebras of the sect.2 in terms of time-variables. Indeed, for
convenience of taking the continuum limit, the time variables should be
redefined $(i.e$. the integrable flows of the previous section are not suitable
in the continuum limit). In other words, this is the question what is the
proper reduction, or what combinations of the ``integrable" times should be
eliminated.

To begin with, we consider the simplest non-trivial case of $p=3$. Then
introducing the scalar fields
\beq\label{d1}
\partial \phi ^{(1)}(z) = \sum  _k kt^{(1)}_kz^{k-1} + \sum  _k
{\partial \over \partial t^{(2)}_k} z^{-k-1},
\eeq
\beq\label{d2}
\partial \phi ^{(2)}(z) = \sum  _k kt^{(2)}_kz^{k-1} + \sum  _k
{\partial \over \partial t^{(1)}_k} z^{-k-1},
\eeq
with  $t^{(1)}_k= (i\bar t_k+t_k)/2\sqrt2$,
$t^{(2)}_k= (i\bar t_k-t_k)/2\sqrt2$, one
obtains the expressions:
\beq\label{d3}
W^{(2)}(z) = {1\over2}\partial \phi ^{(1)}(z)\partial \phi ^{(2)}(z),
\eeq
\beq\label{d4}
W^{(3)}(z) = {1\over3\sqrt3}\sum  _i (\partial \phi ^{(i)}(z))^3.
\eeq
instead of (\ref{23}) and (\ref{24}).

This choice of basis in the Cartan plane is adequate to the continuum limit of
the system under consideration, as the latter one is described by completely
analogous expressions \cite{FKN91a}.
Now let us describe this basis in more invariant
terms and find the generalization to arbitrary $p.$

Comparing (\ref{d4}) with (\ref{26}), we can conclude that  $\partial
\phi ^{(i)}\equiv
\bbeta _i\partial \bphi $  corresponds to the basis
\beq\label{d5}
\bbeta _{1,2} = {1\over 2}(\sqrt{3}\bmu _2 \pm  i\balpha _2).
\eeq
This basis has the properties
\beq\label{d6}
\bbeta _1\cdot \bbeta _2 = 1,  \ \ \  \bbeta _1\cdot \bbeta _1 = 0,  \ \ \
\bbeta _2\cdot \bbeta _2 = 0.
\eeq

Now it is rather evident how this basis should look in the case of
general $p$. Due to \cite{FKN91a} we can guess what is the choice of the proper
scalar fields:
\beq\label{d7}
\partial \phi ^{(i)}(z) = \sum  _k kt^{(i)}_kz^{k-1} + \sum  _k
{\partial \over \partial t^{(p-i)}_k} z^{-k-1}.
\eeq
This choice certainly corresponds to the basis with defining property (it can
be observed immediately from the relations (\ref{d7}) and (\ref{17})):
\beq\label{d8}
\bbeta _i\cdot \bbeta _j = \delta _{p,i+j},
\eeq
the proper choice of the Hamiltonians in (\ref{16}) being
\beq\label{d9}
H = \sum _{i,k}t^{(i)}_k\bbeta _i\cdot \bJ_k,
\eeq
what determines new times adequate to the continuum limit.

Let us construct the basis (\ref{d8}) in a manifest way.
To begin with, we define a set of vectors $\{\mu _i\}$ with the property:
\beq\label{d10}
\bmu _i\cdot \bmu _j = \delta _{ij} - {1\over p}, \ \ \  \sum  _i \mu _i = 0.
\eeq
The $W^{(n)}$-algebra can be written in this basis as follows \cite{FL88}:
\beq\label{d11}
W^{(n)} = (-)^{n+1} \sum _{1\leq j_1<\ldots<j_k\leq p}~ \prod ^n_{m=1}
(\bmu _{j_m}\cdot \partial \bphi ),  \ \ \   n=1,2,\ldots,p.
\eeq
Now the basis (\ref{d8}) can be constructed from (\ref{d10})
by diagonalization of the
following cyclic permutation \cite{FKN91a,FKN91b}:
\beq\label{d12}
\bmu _i \rightarrow  \bmu _{i+1}, \ \ \  \bmu _p \rightarrow  \bmu _1\ \ \
i=1,\ldots,p-1 .
\eeq
This transformation has $\{\bbeta _i\}$ as its eigenvectors, their manifest
expressions being of the form:
\beq\label{d13}
\bbeta _k = {1\over \sqrt p} \sum ^p_{j=1} \exp \{{2\pi i\over p}jk\}
\bmu _j,  \ \ \   k=1,2,\ldots,p-1.
\eeq
It is trivial to check that the properties (\ref{d8}) are indeed satisfied.
One can immediately rewrite the corresponding $W$-generators in the basis of
$\bbeta _i$'s. After all, one obtain the expressions similar to the continuum
$W$-generators \cite{FKN91a,FKN91b},
but with the scalar fields defined as in (\ref{d7})
and without
the ``anomaly" corrections appearing in the continuum case due to the twisted
boundary conditions. These corrections can be correctly reproduced by taking
the $p$-th root of the partition function as well as simultaneously doing the
reduction (see subsects.4.3-4.5).

Thus, the proposed procedure allows one to take the continuum limit immediately
transforming the scalar fields as elementary building blocks.

\subsection{The  two-\mm example}

In this section we consider the simplest example of the $p=3$ CMM.

{\em From $\{t\}$- to $\{T\}$-variables.\ }
To describe this change of variables we shall use the scalar-field formalism.

Consider the set of scalar fields (\ref{d1}),(\ref{d2}) and perform the
reduction
\beq\label{e1}
t^{(i)}_{3k+1}=t^{(i)}_{3k+2}=0,\ i=1,2,\ k=0,1,\ldots\ \ .
\eeq
Then
\bea\label{e2}
\partial\varphi^{(1)}(u)=\sum_{k\geq0}g_k^{(1)}u^{3k-1}+
\sum_{k\geq1}{\partial\over\partial t^{(2)}_{3k}}u^{-3k-1},\nn\\
\partial\varphi^{(2)}(u)=\sum_{k\geq0}g_k^{(2)}u^{3k-1}+
\sum_{k\geq1}{\partial\over\partial t^{(1)}_{3k}}u^{-3k-1},
\eea
and $g^{(i)}_k=kt^{(i)}_{3k},\ k=1,2,\ldots$, $g^{(i)}_0=N^{(i)}$  were
introduced in the subsect.4.2. $N^{(1,2)}$ are ``the
dimensions of the matrices'' used in our two-matrix model.
Then we put
\beq\label{e3}
{1\over\tilde\tau}\partial\Phi^{(i)}(z)\tilde\tau =
a u^{i-2}{1\over\tau^{\rm red}}\partial\varphi^{(i)}(u)\tau^{\rm red},~~\
u^3=1+az,\ i=1,2,
\eeq
where $\Phi^{(i)}(z)$ are the scalar fields of the continuum model
\cite{FKN91a}
\bea\label{e4}
& \partial\Phi^{(1)}(z)=\sum_{k\geq0}\left(\left(k+{1\over3}\right)T_{3k+1}
z^{k-{2\over3}}+
{\partial\over\partial \tilde T_{3k+2}}z^{-k-{5\over3}}\right),\nn\\
& \partial\Phi^{(2)}(z)=\sum_{k\geq0}\left(\left(k+{2\over3}\right)T_{3k+2}
z^{k-{1\over3}}+
{\partial\over\partial \tilde T_{3k+1}}z^{-k-{4\over3}}\right),
\eea
and
\beq\label{ee44}
 T_{3k+i}=\tilde T_{3k+i}+a{k\over k+i/3}\tilde T_{3(k-1)+i},\
i=1,2 .
\eeq
The relation (\ref{e3}) gives rise to
the following Kazakov-like change of the time
variables
\beq\label{e5}
g^{(i)}_m=\sum_{n\geq m}{(-)^{n-m}\Gamma\left(n+1+{i\over3}\right)
a^{-n-{i\over3}}\over(n-m)!\Gamma\left(m+{i\over3}\right)}T_{3n+i},\
i=1,2
\eeq
and equations
(\ref{e5}) can be also continued to the zero
discrete times $g_0^{(i)}$.    It follows from (\ref{e3}) that   the
derivatives with respect to  $t_{3k}^{(i)},\ i=1,2$ transform  as
\beq\label{e6}
{\partial\over\partial t^{(i)}_{3k}}=\sum_{n=0
}^{k-1}{\Gamma\left(k+{i\over3}\right)
a^{n+{i\over3}}\over(k-n-1)!\Gamma\left(n+1+{i\over3}\right)}{\partial
\over\partial \tilde T_{3n+i}},\ i=1,2
\eeq
To cancel the terms like $T_{3k+2}z^{-k-{4\over3}}$ and
$T_{3k+1}z^{-k-{5\over3}}$ with $k\geq0$ at the left hand sides of
(\ref{e3}) we have to rescale the partition function of the
continuum model $\tilde \tau$ by means of an exponent of  some
quadratic form
\beq\label{e7}
\tilde\tau=\exp\left(-\sum_{m,n\geq0}A_{mn}\tilde T_{3m+2}
\tilde T_{3n+1} \right)\tau^{\rm red},
\eeq
where
$A_{nm}$ has the form
\beq\label{e16}
A_{nm}={\Gamma\left(n+{5\over3}\right)\Gamma\left(m+{4\over3}\right)\over
\Gamma\left({2\over3}\right)\Gamma\left({1\over3}\right)}
{(-)^{n+m}a^{-n-m-1}\over n!m!(n+m+1)(n+m+2)}.
\eeq
\medskip

{\em Constraint algebras of the two-matrix model.}\ \
It was suggested in \cite{FKN91a} that the  continuum
``two-matrix'' model  {\em possesses}  the ${\cal W}^{(3)}$
 (including Virasoro) symmetry, where
Virasoro generators ${\cal L}_n$ and generators of the ${\cal W}^{(3)}$
algebra are constructed from the scalar fields (\ref{e4})
in the following way
\beq\label{e17}
{\cal T}(z)={1\over2}
{:}\partial\Phi^{(1)}(z)\partial\Phi^{(2)}(z){:}-{1\over9z^2}
=\sum{{\cal L}_n\over
z^{n+2}},
\eeq
\beq\label{e18}
W^{(3)}(z)={1\over3\sqrt3}
\left({:}\left(\partial\Phi^{(1)}(z)\right)^3{:}+
{:}\left(\partial\Phi^{(2)}(z)\right)^3{:}\right)
=\sum{{\cal W}_n^{(3)}\over
z^{n+3}}.
\eeq

One can easily derive  a
relation between the generators $\tilde{\cal L}_n$
and the corresponding generators $L_{3n}^{\rm red}$, associated with reduction
(\ref{e1}):
\beq\label{e19}
{1\over\tilde\tau}\tilde{\cal L}_n\tilde\tau=
a^{-n}\sum_{p=0}^{n+1}C_p^{n+1}(-)^{n+1-p}
{L_{3p}^{\rm red}\tau^{\rm red}\over\tau^{\rm red}},\
\ \ n\geq-1,
\eeq
where Virasoro generators $L_{3n}^{\rm red}$ are defined by the same formula
(\ref{d3}), only with $\phi$ substituted by ``reduced'' fields $\varphi$:
\beq\label{e24}
T(u)={1\over2}{:}\partial\varphi^{(1)}(u)\partial\varphi^{(2)}(u){:}=
\sum_{n}u^{-3n-2}L_{3n}^{\rm red},
\eeq
and
\bea\label{e20} \tilde{\cal
L}_{-1}=&\sum_{k\geq1}\left(\left(k+{1\over3}\right)T_{3k+1}
{\partial\over\partial \tilde T_{3k-2}}+
\left(k+{2\over3}\right)T_{3k+2}
{\partial\over\partial \tilde T_{3k-1}}\right)~+~{2\over9}T_1T_2,\nn
\\
\tilde{\cal L}_{0}=&\sum_{k\geq0}\left(\left(k+{1\over3}\right)T_{3k+1}
{\partial\over\partial \tilde T_{3k+1}}+
\left(k+{2\over3}\right)T_{3k+2}
{\partial\over\partial \tilde T_{3k+2}}\right), \nn\\
\tilde{\cal L}_n=&\sum_{k-m=-n}\left(\left(k+{1\over3}\right)T_{3k+1}
{\partial\over\partial \tilde T_{3m+1}}+
\left(k+{2\over3}\right)T_{3k+2}
{\partial\over\partial \tilde T_{3m+2}}\right)  \nn\\
&+\sum_{m+k=n-1}{\partial\over\partial \tilde T_{3k+2}}
{\partial\over\partial \tilde T_{3m+1}} +{(-)^n\over9a^n},\ \ \
n\geq1.
\eea
Eq. (\ref{e19}) is a direct consequence of the relation
\bea\label{e200}
{1\over\tilde\tau}{\cal T}(z)\tilde\tau &=
{1\over\tilde\tau}\partial\Phi^{(1)}(z)
\partial\Phi^{(2)}(z)\tilde\tau={1\over\tau^{\rm red}}\partial\varphi^{(1)}(u)
\partial\varphi^{(u)}(u)\tau^{\rm red} = {1\over\tau^{\rm red}}T(u)\tau^{\rm
red},
\nn\\
u^3&=1+az
\eea
(compare with (\ref{e3})). As in the 1-matrix case, the
 generators (\ref{e20}) differ from the continuum generators ${\cal
L}_n$ of \cite{FKN91a} by singular $c$-number terms. Instead, again
$L_{3n}^{\rm red}$ do not exactly annihilate $\tau^{\rm red}$. We assume that
again
these two effects cancel each other, provided eq.(\ref{e19}) for $\tilde\tau$
is rewritten in terms of the $cubic$ root  $\sqrt[3]{\tilde\tau}$.
In orther words, doing accurately the reduction procedure in the Virasoro
constraints of the discrete two-\mm one can rewrite
(\ref{e19}) in the form
\beq\label{ee19}
{1\over\sqrt[3]{\tilde\tau}}{\cal L}^{\rm cont}_n\sqrt[3]{\tilde\tau}=
a^{-n}\sum_{p=0}^{n+1}C_p^{n+1}(-)^{n+1-p}\left.{L_{3p}\tau\over\tau}
\left( 1+{\cal O}(a) \right)\right\vert_{t_{3k+i}=0;\  i=1,2},\
\ \ n\geq-1,
\eeq
Thus, we conclude that
the Virasoro constraints of the continuum \mm are indeed implied by  the
corresponding Virasoro constraints of the discrete $conformal$ two-matrix
model.
\medskip

It follows from (\ref{e18}) that generators of the $W^{(3)}$ symmetry
for the continuous model can be written in the form
\bea\label{e30}
\tilde{\cal W}_n^{(3)}=
&3\sum_{k,m\geq0}
\left(k+{1\over3}\right)\left(m+{1\over3}\right)T_{3k+1}T_{3m+1}
{\partial\over\partial\tilde T_{3(k+m+n)+1}}\nn\\
+&3\sum_{k,m\geq0}
\left(k+{2\over3}\right)\left(m+{2\over3}\right)T_{3k+2}T_{3m+2}
{\partial\over\partial\tilde T_{3(k+m+n-1)+2}}\nn\\
+&3\sum_{m,p\geq0}
\left(m+p-n+{4\over3}\right)T_{3(m+p-n+1)+1}
{\partial\over\partial\tilde T_{3m+2}}
{\partial\over\partial\tilde T_{3p+2}}\nn\\
+&3\sum_{m,p\geq0}
\left(m+p-n+{2\over3}\right)T_{3(m+p-n)+2}
{\partial\over\partial\tilde T_{3m+1}}
{\partial\over\partial\tilde T_{3p+1}}\nn\\
+&
\sum_{k,m\geq0}
{\partial\over\partial\tilde T_{3k+2}}
{\partial\over\partial\tilde T_{3m+2}}
{\partial\over\partial\tilde T_{3(n-k-m)-4}}+
{\partial\over\partial\tilde T_{3k+1}}
{\partial\over\partial\tilde T_{3m+1}}
{\partial\over\partial\tilde T_{3(n-k-m)-2}},
\eea
where $n\geq-2$ and the terms with negative values of the indices
should be omitted. The generators ${\cal W}_{-2}^{(3)}$  and
${\cal W}_{-1}^{(3)}$ (but not the ${\cal W}_0^{(3)}$) have additional
terms, cubic in times:
\beq\label{e31}
{\cal W}_{-1}^{(3)}={1\over27}T_1^3+\cdots\ \ \hbox{and}\ \
{\cal W}_{-2}^{(3)}={8\over27}T_2^3+{4\over9}T_1^2T_4+\cdots\ .
\eeq

Similar to the Virasoro case   one can show that
the relation between the generators of the $W^{(3)}$ symmetry
for the discrete  model and the generators $\tilde{\cal W}_n^{(3)}$
is
\beq\label{e32} {1\over\tilde\tau}\tilde{\cal
W}_p^{(3)}\tilde\tau=
a^{-p}\sum_{n=0}^{p+2}C_{p+2}^n(-)^{p-n}
\frac{W^{(3)\ {\rm red}}_{3n}\tau^{\rm red}}{\tau^{\rm red}},\
\ p\geq 2
\eeq
and follows from the identity \beq\label{e33}
\left({\partial\Phi^{(1)}(z)\tilde\tau\over\tilde\tau}\right)^3
+\left({\partial\Phi^{(2)}(z)\tilde\tau\over\tilde\tau}\right)^3=
a^3\left[{1\over
u^3}\left({\partial\varphi^{(1)}(u)\tau^{\rm red}\over\tau^{\rm red}}\right)^3
+\left({\partial\varphi^{(2)}(u)\tau^{\rm red}\over\tau^{\rm
red}}\right)^3\right],
\eeq
where the generators $W_n^{(3)\ {\rm red}}$ of the discrete model are defined
after the reduction (\ref{e1}) by the relation
\beq\label{e34}
\left[{1\over
u^3}\left(\partial\varphi^{(1)}(u)\right)^3
+\left(\partial\varphi^{(2)}(u)\right)^3\right]=\sum_{n}u^{-3n-3}
W^{(3)\ {\rm red}}_{3n}.
\eeq

After reduction of the discrete model eq.(\ref{e32})
can be rewritten in the form
\beq\label{ee32}
{1\over\sqrt[3]{\tilde\tau}}{\cal
W}_p^{(3)\ {\rm cont}}\sqrt[3]{\tilde\tau}=
a^{-p}\sum_{n=0}^{p+2}C_{p+2}^n(-)^{p-n}\left.\frac{W^{(3)}_{3n}\tau}{\tau}
\right\vert_{t_{3k+i}=0;\  i=1,2},\
\ p\geq 2 ,
\eeq
where          ${\cal W}_p^{(3){\rm cont}}$ are the generators of the
${\cal W}^{(3)}$-symmetry from the paper \cite{FKN91a} (see (\ref{e30}) and
(\ref{e31})).

\subsection{The general case}

It is easy to generalize the example of $p=3$ to general multi-\mms
using the formalism of scalar fields with ${\bf Z}_p$-twisted boundary
conditions. Let us introduce $p-1$ sets of the discrete times
$t^{(i)}_{k}$, $i=1,2,\ldots,p-1$ and $k=0,1,\ldots$\ for the discrete
$(p-1)$-\mm and consider the reduction
\beq\label{e41}
t^{(i)}_{pk+j}=0,\ i,j=1,2,\ldots,p-1,\ k=0,1,\ldots\ \ .
\eeq
Choose  the discrete and continuum scalar fields in the form
\beq\label{e42}
\partial\varphi^{(i)}(u)=\sum_{k\geq0}g_k^{(i)}u^{pk-1}+
\sum_{k\geq1}{\partial\over\partial t^{(p-i)}_{pk}}u^{-pk-1},
\quad g_k^{(i)}=kt^{(i)}_{pk}
\eeq
\beq\label{e44}
\partial\Phi^{(i)}(z)=\sum_{k\geq0}\left\{\left(k+{i\over p}\right)T_{pk+i}
z^{k-{p-i\over p}}+
{\partial\over\partial \tilde T_{pk+p-i}}z^{-k-{2p-i\over  p}}\right\},
\eeq
\beq\label{ee43}
  T_{pk+i}=\tilde T_{pk+i}+a{k\over k+i/p}\tilde T_{p(k-1)+i},\
i=1,2,\ldots,p-1 .
\eeq

Then the equations
\beq\label{e43}
a u^{i-p+1}{1\over\tau^{\rm red}}\partial\varphi^{(i)}(u)\tau^{\rm red}=
{1\over\tilde\tau}\partial\Phi^{(i)}(z)\tilde\tau,\
u^p=1+az,\ i=1,2,\ldots,p-1,
\eeq
generate the Kazakov-like change of the time variables
\beq\label{e45}
g^{(i)}_m=\sum_{n\geq m}{(-)^{n-m}\Gamma\left(n+1+{i\over p}\right)
a^{-n-{i\over p}}\over(n-m)!\Gamma\left(m+{i\over p}\right)}T_{pn+i},\
i=1,2,\ldots,p-1
\eeq
\beq\label{e46}
{\partial\over\partial t^{(i)}_{pk}}=\sum_{n=0
}^{k-1}{\Gamma\left(k+{i\over p}\right)
a^{n+{i\over p}}\over(k-n-1)!\Gamma\left(n+1+{i\over p}\right)}{\partial
\over\partial \tilde T_{pn+i}},\ i=1,2,\ldots,p-1.
\eeq

Using the eq. (\ref{e43}) and considerations similar to
those used in the previous
subsections one can show that            there is relation
between tilded continuum generators $\tilde{\cal W}^{(i)}_n$,
$i=2,\ldots,p$  of the ${\cal W}$-symmetry and reduced generators of
the discrete $W$-symmetry $W_{pk}^{(i){\rm red}}$  of the form
\beq\label{e49}
{1\over\tilde\tau}\tilde{\cal W}_n^{(i)}\tilde\tau=
a^{-n}\sum_{s=0}^{n+i-1} C_s^{n+i-1}(-)^{n+i-1-s}
\frac{W^{(i){\rm red}}_{ps}\tau^{\rm red}}
{\tau^{\rm red}},\ \ \ \  n\geq -i+1,
\eeq
where rescaled $\tau$-function is defined
\beq\label{e47}
\tilde\tau=\exp\left(-{1\over2}\sum_{i=1}^{p-1}\sum_{m,n\geq0}A^{(i)}_{mn}
\tilde T_{pm+i}
\tilde T_{pn+p-i} \right)\tau^{\rm red}\{t_{pk}\}
\eeq
and matrices $A^{(i)}_{nm}$ are determined by
\beq\label{e416}
A^{(i)}_{nm}={\Gamma\left(n+{p+i\over p}\right)\Gamma\left(m+{2p-i\over p}
\right)\over
\Gamma\left({i\over p}\right)\Gamma\left({p-i\over p}\right)}
{(-)^{n+m}a^{-n-m-1}\over n!m!(n+m+1)(n+m+2)},\  i=1,2,\ldots,p-1.
\eeq
This relation again corresponds to the identity
\beq
\frac{W(u)\tau^{\rm red}}{\tau^{\rm red}}=\frac{{\cal
W}(z)\tilde\tau}{\tilde\tau},
\eeq
where $u^p=1+az$.

Performing the proper reduction procedure (\ref{e41}), which eliminates all
but the time-variables of the form $t_{pk}^{(i)}$
(i.e. leaves the $1/p$ fraction
of the entire quantity of variables) we can obtain the relation
\beq\label{ee49}
{1\over\sqrt[p]{\tilde\tau}}{\cal W}_n^{(i)}\sqrt[p]{\tilde\tau}=
a^{-n}\sum_{s=0}^{n+i-1}C_s^{n+i-1}(-)^{n+i-1-s}\left.{W^{(i)}_{ps}
\tau\over\tau}\right\vert_{t_{pk}^{(i)}\neq 0\ \ {\rm only}}
\left( 1+ O(a) \right),\ \ \ n\geq-i+1,
\eeq
where ${\cal W}_n^{(i)}$ is the ${\cal W}$-generators of the paper
\cite{FKN91a}.
Thus, we proved the $W$-invariance of the partition function of the
continuum $p-1$-\mm and found the explicit relation between
its partition function and
corresponding partition function of the discrete $(p-1)$-matrix model.

\subsection{On the reduction of the partition function}

To conclude this section we would like to discuss the problem of reduction
(\ref{e1}) and (\ref{e41}) of the partition function in detail,
with accuracy up to (non-leading) $c$-number contributions and only
after the continuum limit is taken. More
precisely, we reformulate the  condition of a proper reduction in the continuum
limit in order to reduce it to more explicit formulas which can be
immediately checked. As a by-product of our consideration we obtain some
restrictions on the integration contour in the partition function (\ref{21}).

To get some insight, let us consider the simplest case of the Virasoro
constrained Hermitian one-\mm \cite{MMMM91}.
Before the reduction the Virasoro operators read as in
(\ref{b5}). Then their action on $\log \tau$ can be rewritten as
\beq\label{f1}
\left[ \sum_k kt_k {\partial \log \tau \over \partial t_{k+n}} +
\sum_m {\partial ^2 \log \tau \over \partial t_m \partial t_{n-m}}\right] +
\sum_m \left[{\partial \log \tau \over \partial t_m}
{\partial \log \tau \over \partial t_{n-m}}\right] = 0.
\eeq

After the reduction, we obtain
\bea\label{f2}
&\left[ \sum_k 2kt_{2k} {\partial \log \tau ^{\rm red}\over \partial t_{2k+2n}}
+
\sum_m {\partial ^2 \log \tau ^{\rm red} \over
\partial t_{2m} \partial t_{2n-2m}} +
\sum_m {\partial ^2 \log \tau ^{\rm red} \over
\partial t_{2m+1} \partial t_{2n-2m-1}} \right] + \nn\\
&+ \sum_m \left[{\partial \log \tau ^{\rm red} \over \partial t_{2m}}
{\partial \log \tau ^{\rm red} \over \partial t_{2n-2m}}\right] = 0
\eea
under the condition
\beq\label{f3}
\left.{\partial \log \tau ^{\rm red} \over \partial t_{\rm odd}}\right|_{t_{\rm
odd}=0} = 0.
\eeq

The last formula is a direct consequence of the ``Schwinger-Dyson'' equation
induced by the transformation of the reflection $M \to -M$ in (\ref{b4}).
Indeed, due to the invariance of the integration measure under this
transformation one can conclude that the partition function (\ref{b4}) depends
only on a quadratic form of odd times.

Thus, the second derivatives of $\log \tau$ over odd times do not vanish, and
are conjectured to satisfy the relation
\beq\label{f4}
\sum_m {\partial ^2 \log \tau ^{red} \over
\partial t_{2m} \partial t_{2n-2m}} \sim
\sum_m {\partial ^2 \log \tau ^{red} \over
\partial t_{2m+1} \partial t_{2n-2m-1}},
\eeq
where the sign $\sim$ implies that this relation should be correct only
{\it after} taking the continuum limit.
In this case one obtains the final result (cf. (\ref{a16}))
\beq\label{f5}
\left[ \sum_k kt_{2k} {\partial \log \sqrt{\tau ^{red}}
\over \partial t_{2k+2n}} +
\sum_m {\partial ^2 \log \sqrt{\tau ^{red}} \over \partial t_{2m}
\partial t_{2n-2m}}\right] = 0.
\eeq

Thus, it remains to check the correctness of the relation (\ref{f4}).
To do this,
one should use the manifest equations of integrable (Toda chain) hierarchy,
and after direct but tedious calulations \cite{MMMM91} one obtains the result
different from the relation (\ref{f4}) by $c$-number terms which are singular
in the limit $a\to 0$ and just cancell corresponding items in (\ref{b6})
(this is certainly correct only after taking the continuum limit).

All this (rather rough) consideration can be easily generalized to the
$p$-\mm case.
In this case one should try to use all ${\cal W}_n^{(i)}$-constraints with
$2\le i\le p$. Thus, the second derivatives should be replaced by higher
order derivatives, and one obtain a series of equations like (\ref{f1}). It is
the matter of trivial calculation to check that these equations really give
rise to the
proper constraints satisfied by $\sqrt[p]{\tau}$ (cf. (\ref{ee19}),
(\ref{32}) and (\ref{ee49}))
provided by the two sets of the relations like (\ref{f3}) and (\ref{f4}) .

Namely, the analog of the relation (\ref{f3}) in the $p$-\mm case is the
cancellation of all derivatives with incorrect gradation, i.e. with the
gradation non-equal to zero by modulo $p$. The other relation (\ref{f4})
should be replaced now by the conditions of the equality (in the continuum
limit) of {\it all} possible terms with the same {\it correct}
gradation. In the simplest case of $p=3$ these are
\bea\label{f6}
&\sum_m {\partial ^2 \log \tau ^{red} \over
\partial t_{3m+1} \partial t_{3n-3m-1}} \sim
\sum_m {\partial ^2 \log \tau ^{red} \over
\partial t_{3m} \partial t_{3n-3m}},\nn\\
&\sum_{m,k} {\partial ^3 \log \tau ^{red} \over
\partial t_{3m+1} \partial t_{3n-3(m+k)+1} \partial t_{3k-2}} \sim
\sum_{m,k} {\partial ^3 \log \tau ^{red} \over
\partial t_{3m+2} \partial t_{3n-3(m+k)+2} \partial t_{3k-4}} \sim\nn\\
&\sim \sum_{m,k} {\partial ^3 \log \tau ^{red} \over
\partial t_{3m+2} \partial t_{3n-3(m+k)+1} \partial t_{3k}} \sim
\sum_{m,k} {\partial ^3 \log \tau ^{red} \over
\partial t_{3m} \partial t_{3n-3(m+k)} \partial t_{3k}}.
\eea

Again, this second condition is correct modulo some singular in the limit of
$a\to 0$ terms, which appear only in the case of even $p$. Unfortunately,
we do not know the way to prove this statement without using the integrable
equations, what is very hard to proceed in the case of higher $p$.

On the other hand, the cancellation of derivatives with
incorrect gradation can be trivially derived from the ``Schwinger-Dyson''
equations given rise by the transformations
$M\to \exp\left\{{2\pi ki\over p}\right\}M$ ($0<k<p$) of the integration
variable in the
corresponding matrix integral, the integration measure being assumed to be
invariant. In its turn, it implies that the integration contour, instead of
real line, should be chosen as a set of rays beginning in the origin
of the co-ordinate system with the angles between them being integer times
$2\pi \over p$. This rather fancy choice of the integration contour
is certainly necessary to preserve ${\bf }Z_p$-invariance of $p$-\mm system.

\section{Conclusion}

In this paper we proposed a new point of view on the discrete matrix
formulation of the unified theory of $2d$ gravity coupled to minimal
series of $c<1$ matter. As alternative to
conventional discrete multi-matrix models which are extremely hard to
investigate by imposing some differential equations  (Ward identities etc.)
on matrix path integral, we introduced another class of models
which satisfy ``conformal" constraints by definition and actually possess much
richer integrable structure, essentially as rich as the one arising in
continuum limit. This is also up to now the only case when
continuum double scaling interpolating limit (of multi-matrix models)
can be performed honestly, reproducing the results of \cite{FKN91a}.

If matrix models are considered as certain solutions to integrable theories,
the conformal multi-matrix models satisfy the equations of multi-component KP
hierarchy. More precisely, they correspond to particular reduction of
multi-component KP hierarchy, the reduction of AKNS type, which is further
constrained to be consistent with discrete $W$-constraints and/or discrete
string equation. We
found a determinant representation for a certain subclass of such solutions,
which can be also considered as a generalization of those corresponding to
orthogonal polynomials,
leading to {\it forced} hierarchies. All this must shed light on the origin of
the integrability in $2d$ gravity and the work only started in this direction.

So far there was no real progress in taking the continuum limit of conventional
multi-matrix models, and most of the facts were rather introduced axiomatically
in this case. In contrast to this, CMM do have a nice continuum limit, which
can be described in the same terms as the continuum limit of 1-matrix case.
This shows that the fact of discrete $W$-invariance or, put
differently, the proper choice of a particular representative within the
universality class, is crucial to expose the origin of string equation
of $2d$ gravity theory in a manifest form.

Certainly there must be a deep connection between CMM and unified description
of the continuum theory via Generalized Kontsevich Model (GKM)
\cite{KMMMZ91a,KMMMZ91b,KMMM92a,KMMM92b,Mar92}.
We know \cite{KMMM92a} that there exists also a reformulation of a
discrete Hermitian 1-matrix model in terms of GKM, the integrability being
established using formulas very close to those of the
sect.3. It is possible also
to take the continuum limit of 1-matrix case in ``internal GKM terms",
what is much easier than by technique developed in \cite{MMMM91} and sect.4
above. All these facts implies that there should be a sort of reformulation of
the CMM in GKM-like terms.

There is another still unresolved question on interpretation of conformal
fields appearing in the above formalism and their relations to the conformal
fields of Polyakov's formulation of 2d gravity. Of course, there shouldn't be
any direct correspondence, but there can be a kind of duality
between ``world-sheet"
and ``spectral" Riemann surfaces. The key role of $W$-constraints found in the
paper and the presence of multi-component scalar fields might also have a
meaning in the framework of $W$-gravity theory \cite{GLM91,MMMO92}.

We hope to return to all these problems elsewhere.

\section{Acknowledgments}

We are grateful for stimulating discussions to C.Bachas, B.Gato-Rivera,
A.Orlov and A.Zabrodin. One of us (A.Mir.) is indepted to E.Antonov for
the kind hospitality in S.-Petersburg where a piece of this paper was done.

\end{document}